\documentstyle[preprint,aps]{revtex}
\begin{document}
\draft
\preprint{IISc-PHYCMT : 94/15 }
\title{Nonequilibrium Phase Transitions in a Driven Sandpile
Model}
\author{Sujan K. Dhar,
Rahul Pandit\cite{byjnc}, and Sriram Ramaswamy$^{\,1}$ \cite{byjnc}}
\address{Department of Physics, and ${\,^1}$Centre for
Theoretical Studies, Indian Institute of Science,\\
Bangalore - 560 012, India}
\date{\today}
\maketitle
\begin{abstract}
We construct a driven sandpile slope model and study it by
numerical simulations in one dimension. The model is specified
by a threshold slope $\sigma_c\/$, a parameter $\alpha\/$,
governing the local current-slope relation (beyond threshold),
and $j_{\rm in}$, the mean input current of sand. A
nonequilibrium phase diagram is obtained in the $\alpha\, -\,
j_{\rm in}\/$ plane.  We find an infinity of phases,
characterized by different mean slopes and separated by
continuous or first-order boundaries, some of which we obtain
analytically. Extensions to two dimensions are discussed.
\end{abstract}

\narrowtext
The statistical mechanics of nonequilibrium steady states is a
subject of growing general interest.  Phase transitions between
such states are by no means as well understood as their
equilibrium counterparts. Some insight has been gained into this
problem by the study of simple driven lattice models
\cite{ddlg,toom}. Sandpile models, which introduced the notion
of {\em self-organized criticality} (SOC) \cite{btw} as a
general explanation for the wide occurrence of power laws in
nature, are a natural setting in which to study phase
transitions far from equilibrium. Surprisingly, to our
knowledge, this has not been attempted.  We construct a driven
sandpile model and show that it exhibits, in spite of its
simplicity, a rich phase diagram, thus making it a good
laboratory for the study of nonequilibrium phenomena.  In
particular, our model exhibits continuous transitions from
pinned or threshold-dominated states to unpinned states; these
are reminiscent of dynamical phase transitions in more complex
systems such as sliding charge-density waves (CDW's)\cite{onu}
and pinned flux-lattices
\cite{flux}.

Earlier studies of sandpile models have concentrated on SOC,
either in steadily flowing sandpiles \cite{hka,carl} or at the
angle of repose \cite{kad}. Much of the work has been on {\em
critical-height} models, in which the update rule depends only
on the {\em height} at each site; {\em critical-slope} models
(CSM) \cite{kad} have been studied to a lesser extent. In this
Letter, we present a comprehensive study of a simple CSM in
which the current-slope relation for slopes exceeding a
threshold is controlled by a parameter $\alpha\/$. We monitor
the steady states of our model as a function of $\alpha\/$ and
the mean input current $j_{\rm in}\/$ and find the rich
nonequilibrium phase diagram (Fig. 1): It shows that there are
many phases characterized by the average slope $\sigma_{\rm
av}\/$ of the sandpile. In the {\em repose} phase, $\sigma_{\rm
av} = \sigma_c\/$, the slope at the angle of repose.  As $j_{\rm
in}\/$ is increased at fixed $\alpha\/$, the repose phase
undergoes a continuous transition (solid line) to an unpinned
super-repose phase at which $\sigma_{\rm av} - \sigma_c$ rises
continuously from zero with an exponent $\beta \simeq 0.5\/$
(Fig. 2). This continuous transition is followed by a series
(which we argue is infinite) of similar continuous transitions.
The repose phase lies between two first-order boundaries: one at
low $\alpha$ to a pinned super-repose ($\sigma_{\rm av} >
\sigma_c$) region, the other at large $\alpha$, to a sub-repose
phase ($\sigma_{\rm av} < \sigma_c$). The lower one of these
first-order lines meets the continuous line at $j_{\rm in} =
0.5\/$ at a multicritical point. The pinned super-repose region
contains an infinity of phases, separated by first-order lines
parallel to the $j_{\rm in}$ axis (Fig. 1); $\sigma_{\rm av}$
jumps at these boundaries. We also monitor height profiles, the
equal-time height correlation function and the associated
correlation length, and current autocorrelations and the
associated correlation time.  Both the correlation length and
time (Fig. 3) diverge at the continuous transitions in Fig. 1.
We show that our main results can be understood on the basis of
a mean-field theory and a mode-softening argument.

In our model, integers $h_{i}\/$ specify the heights of columns
of sand at the sites $i\/$ of a one-dimensional chain $(1 \leq i
\leq N)\/$. The stability of the column at a site is determined
by a threshold condition which mimics the angle of repose for a
real sandpile: When the height difference between a site $i\/$
and its right neighbor $(i+1)\/$ (i.e., the local slope
$\sigma_i = h_{i} - h_{i+1}\/$) exceeds $\sigma_c\/$, some sand
topples to the right neighbor, and $h_i\/$ is updated via
\begin{equation}
h_i \rightarrow h_i -  j_i\, ,\quad h_{i+1} \rightarrow h_{i+1}
+  j_i \; ,\label{eq:trans}
\end{equation}
\begin{equation}
j_i = {\cal N}\bigl( \alpha \times \sigma_i \bigr)
\Theta(\sigma_i-\sigma_c)\, . \label{eq:urule}
\end{equation}
${\cal N}(x)\/$ is the integer nearest to $x\/$, $\Theta\/$ the
step function, and $\alpha\/$ a real number.  This part of our
update rule conserves the number of particles locally and yields
a local current which increases with the local slope, thus
preventing an unbounded buildup of particles in the pile
\cite{foot1}.  The parameter $\alpha\/$ controls
the current-slope relation for slopes exceeding $\sigma_c\/$.
We restrict our study to $\alpha > 0\/$, since $\alpha \leq 0\/$
yields unphysical runaway behavior; the upper bound for
$\alpha\/$ is chosen to limit the region we explore.  The mean
input current of sand particles $j_{\rm in}\/$ is another
control parameter.  At each time step, we add $m\/$ particles to
a randomly chosen site with probability $p\/$, so $j_{\rm in} =
p \times m\/$.  We set $m = 10\/$ for specificity (our results
do not depend on this choice) and cover the range $0.001 \leq p
\leq 0.15\/$, so $0.01
\leq j_{\rm in} \leq 1.5\/$. This addition of particles violates
local particle conservation; and, for such an addition rate, the
mean input current and the noise amplitude {\em per site} vanish
as $N \rightarrow \infty$.  Particles are allowed to leave our
system through the right, but not the left $(i = 1)\/$,
boundary: any particle that reaches the $N^{th}\/$ site is
removed immediately.  Our boundary conditions and update rules
(1-2) clearly pick a direction for the current (from left to
right).

We use initial conditions in which $h_i\/ = \sigma_c (N- i) +
\delta_i\/$, where $\delta_i\/$ is an integer that assumes the
values $0, \; \pm1\/$ with equal probability. We update all
sites simultaneously and allow the system to evolve till it
attains a steady state, i.e., when $j_{\rm in} = j_{\rm out},\/$
the average output current (the average number of particles
dropping out from the open boundary per unit time). In practice,
we say that the steady state has been achieved when these two
currents are within $1\%\/$ of each other.  Once the steady
state is obtained \cite{foot2}, we accumulate data for $10^5 -
10^6\/$ updates per site (UPS). Data for averages are stored
after every $50\/$ UPS.  We also average our data over $10 -
50\/$ different initial conditions.  To minimize boundary
effects we ignore a few sites (three or more if necessary) near
each boundary while computing all averages.

The nonequilibrium steady state of our model can be
characterized by the mean slope $\sigma_{\rm av}$ (the local
slope $\sigma_i\/$ averaged over $i\/$ and many time steps),
which is the order parameter for our model. We also monitor the
mean current $j_{\rm av}\/$, the output current $j_{\rm out}\/$,
the equal-time height correlation function $C_{hh}(r) =
\bigl<\bigl<
\bigl[h_{i}^{t} - \langle h_{i}\rangle_{t}\bigr]
\bigl[h_{i+r}^{t} - \langle
h_{i+r}\rangle_{t}\bigr]\bigl>_{i}\bigl>_{t}$, and the output
current autocorrelation function $C_{jj}(\tau) =
\bigl<\bigl[j_{\rm out}^{t} - \langle j_{\rm out}\rangle_{t}\bigr]
\bigl[j_{\rm out}^{t+\tau} - \langle j_{\rm out}\rangle_{t}\bigr]
\bigr>_{t}$, where $\langle\cdots\rangle_{i}\/$ and
$\langle\cdots\rangle_{t}\/$ denote, respectively, averages over
$i\/$ and time $t\/$.

In Fig. 2 the asymptotes (dashed lines) $\sigma_{\rm av} =
\sigma_c\/$ and $\sigma_{\rm  av} = j_{\rm in}/(2 \alpha)\/$
indicate the behavior of $\sigma_{\rm av}$ at very low ($ <
0.5\/$) and very large ($\gg 0.5\/$) $j_{\rm in}\/$,
respectively.  The full curve (for $\alpha=0.1\/$ and $ N =
128\/$ in Fig. 2) shows that the approach to these limits is
nontrivial: As $j_{\rm in}\/$ increases there are successive
onsets, indicated by arrows. There might well be an infinity of
such onsets (see below), but they become hard to resolve
numerically at large $j_{\rm in}\/$. The inset shows a
finite-size scaling plot at the first of these onsets.  For
$j_{\rm in} \lesssim 0.5 \; ,\sigma_{\rm av} = \sigma_c = 10\/$,
i.e., we have threshold-dominated behavior at low $j_{\rm
in}\/$. The asymptotic behaviors can be understood via the
mean-field theory presented below.

The evolution equation for $h_i^t\/$ is
\begin{equation}
h_{i}^{t+1} - h_{i}^{t} = -j_{i}^{t} + j_{i-1}^{t} +
\eta_i^t \; , \label{eq:cont}
\end{equation}
where the noise $\eta_i^t\/$ has a mean $j_{\rm in}/N\/$ that
accounts for the addition of particles.  If we average over this
noise, then, in the steady state, we get $j_{i} - j_{i-1} =
j_{\rm in}/N\/$. If we impose the boundary condition $j_{i=N} =
j_{\rm in}\/$, we get $j_{i} = i j_{\rm in}/N\/$ and thence a
mean current $j_{\rm av} = \bigl<\bigl<
j_{i}\bigr>_{i}\bigr>_{t} =
\frac{N+1}{2N} j_{\rm in}\/ \approx j_{\rm in}/2\/$ for $N
\rightarrow \infty$.  We use these exact results to check
our simulation. Our mean-field theory assumes that
\begin{equation}
\bigl<\bigl<{\cal N}(\alpha \sigma_i) \Theta (\sigma_i -
\sigma_c)\bigr>_{i}\bigr>_{t} \simeq \bigl<\bigl<{\cal
N}(\alpha\sigma_i)\bigr>_{i}\bigr>_{t} \;
\bigl<\bigl<\Theta(\sigma_i - \sigma_c)\bigr>_{i}\bigr>_{t} \; .
\end{equation}
For large $j_{\rm in}\/$, most $\sigma_i >
\sigma_c\/$, so we further assume $\bigl<\bigl<\Theta(\sigma_i -
\sigma_c)\bigr>_{i}\bigr>_{t} \simeq 1\/$.
If its argument is large, the discontinuities of ${\cal
N} (\alpha\sigma_i)$ are small relative to $\alpha\sigma_i$,
so, on averaging Eq. (\ref{eq:urule}), we make the
approximation $j_{\rm  av} = \bigl<\bigl<{\cal N}(\alpha
\sigma_i)\bigr>_{i}\bigr>_{t} = \alpha \sigma_{\rm  av}$,
whence $\sigma_{\rm av} \approx j_{\rm in}/(2\alpha)\/$, the
large-$j_{\rm in}\/$ asymptote of Fig. 2.  Given these
approximations, Eq.(\ref{eq:cont}) yields a discrete diffusion
equation for $h_{i}^{t}\/$ with a spatially uniform source
$j_{\rm in}/N\/$. If we solve this with our boundary conditions
we get parabolic height profiles for large $j_{\rm in}\/$, which
we also find in our simulations \cite{lop}.

The the current-slope relation Eq. (\ref{eq:urule}), shows that
(Fig. 4), for a uniform profile with slope $\sigma$, no current
flows if $\sigma \leq \sigma_c$. For $\sigma >
\sigma_c\/$, the current grows with the slope in discrete
steps, which reflect the ${\cal N}$ function in Eq.
(\ref{eq:urule}). As we increase $\alpha\/$, the width of the
step at $j_i = 1\/$ shrinks till it becomes a single point at
$\alpha = 0.125\/$, the value at which the continuous transition
at $j_{\rm in} = 0.5\/$ (Fig. 1) terminates.  This can be
understood as follows: If we turn Fig.  4 on its side, we see
that the slope is pinned at $\sigma_c$ below some threshold
value of $j_i\/$. Beyond this threshold, the slope rises sharply
before it saturates at another value of $\sigma\/$. The onsets
in Fig. 2 are just the sharp steps of Fig. 4 rounded by our
spatiotemporal average.

The vanishing of a step in Fig. 4 can also be linked with the
termination of the continuous line via a mode-softening
argument. Consider, e.g., the step at $j_i=1\/$ for which the
$11 \leq \sigma_i \leq 14\/$ (Fig. 4 with $\alpha = 0.1\/$).
All slopes $\sigma_i\/$ in this interval are equivalent in the
sense that the sandpile dispenses the same amount of local
current for all of them. Thus the restoring force, in response
to a change of an onsite slope from $\sigma\/$ to $\sigma+1\/$,
vanishes and leads to divergent correlation lengths and
relaxation times (Fig. 3).  Clearly the infinity of steps in
Fig. 4 imply an infinity of onsets in Fig. 2, though the
large-$j_{\rm in}$ onsets are hard to resolve numerically.

The above arguments do not yield the value of $j_{\rm in}\/$ at
onset since our spatiotemporal average shifts the values of the
thresholds in Fig. 4. The actual value of $j_{\rm in}\/$ at the
onset depends on the distribution of slopes in the interior of
the pile. We have done some numerical and analytic calculations
\cite{lop} on a `single-step model'\cite{foot1}, which justify the
occurrence of the transition at $j_{\rm in}=0.5\/$.

For the first onset we calculate the mean slope near $j_{\rm in}
\simeq 0.5\/$ for $\alpha=0.1\/$ and $N=32, 64, 128, 256\,
\mbox{and}\, 512\/$ and from a finite-size-scaling analysis (Fig. 2)
obtain the exponents $\beta = 0.5$ and $\nu = 4.0$, where $J
\equiv |j_{\rm in}-j_c|/j_c$ with $j_c=0.5$, and we use the
scaling form $\delta\sigma = N^{-\beta/\nu} {\cal F}(J^{\nu}
N)$, with $\delta\sigma \equiv \sigma_{\rm av}-\sigma_c\/$. The
exponent $\nu\/$, which we obtain in this way from finite-size
scaling, might not be the same as the actual correlation length
exponent as has been pointed out in the context of sliding CDW's
\cite{onu}.

In the low-$j_{\rm in}$ regime, as $\alpha\/$ increases from 0
to 0.05, $\sigma_{\rm av}\/$ decreases in steps of one. The
values of $\alpha\/$ at which these steps occur yield the
first-order boundaries of Fig. 1.  All these first-order
boundaries end at critical points in the range $1 \lesssim
j_{\rm in} \lesssim 1.2\/$. The overall envelope of the steps in
$\sigma_{\rm av}\/$ can be fit approximately to a form
$\sigma_{\rm av} \sim 1/(2\alpha)\/$. This behavior follows from
the update rule (Eq. \ref{eq:urule}): ${\cal N}(\alpha
\sigma_i) = 0\/$ for $\alpha \sigma_i < \slantfrac{1}{2}\/$, so,
if $\alpha < 1/(2\sigma_c)\/$ (=0.05 for $\sigma_c = 10\/$),
even a slope $\sigma_i > \sigma_c\/$ may become stable; e.g.,
with $\sigma_c=10\/$ and $\alpha=0.04\/$, ${\cal
N}(\alpha\sigma_i)=0\/$ for $\sigma_i = \sigma_c + 1 \,
\mbox{and}\, \sigma_c + 2\/$, so the slope at the effective angle of
repose turns out to be $\sigma_{\rm eff} = \sigma_c+2 = 12\/$.
As $\alpha\/$ decreases, $\sigma_{\rm eff}\/$ also increases in
discrete steps, leading to the first-order lines at low
$\alpha\/$.

We fit the equal-time height correlation function $C_{hh}(r)\/$
to an exponential form \cite{lop} which then yields a
correlation length. Figure 3 shows this correlation length
$\xi_{hh}\/$ versus $j_{\rm in}\/$ for $\alpha = 0.1\/$ at
different values of $N\/$. Clearly $\xi_{hh}\/$ diverges at
$j_{\rm in} \simeq 0.5$ and also at subsequent continuous
transitions. We have not characterized the divergences by an
exponent because $\xi_{hh}\/$ exceeds our system size somewhat
before the transition (so we do not show data near $j_{\rm in} =
0.5\/$.).

The output current autocorrelation function $C_{jj}(\tau)\/$
does not fit an exponential form very well. However, we have
extracted correlation times $\tau_{jj}\/$ from the area under
$C_{jj}(\tau)\/$ (normalized so that $C_{jj}(\tau=0)=1\/$).
Figure 3 shows plots of $\tau_{jj}\/$ versus $j_{\rm in}\/$ for
$\alpha = 0.1\/$, which sharpen with increasing $N\/$. This
sharpening shows up clearly near $j_{\rm in} = 0.5\/$ and leads
to an increasing trend in $\tau_{jj}\/$ near the next onset
($j_{\rm in} \simeq 1.1\/$).

It is generally believed that nonequilibrium phase transitions
cannot occur in stochastic, one-dimensional models with
short-ranged interactions \cite{lig}. We believe the transitions
in our model occur because the noise amplitude per site vanishes
as $N\rightarrow \infty\/$. This is why our one-dimensional
model with short-range interactions exhibits phase transitions.
We have checked that these phase transitions get rounded if
there is a finite noise amplitude {\em per site} as $N
\rightarrow \infty$. We have also performed simulations on
two-dimensional versions of our model. Our preliminary results
indicate that, with low noise and a variety of boundary
conditions, such two-dimensional models show the same
transitions as the one-dimensional model discussed here. In the
steady state, the two-dimensional system behaves like uncoupled
one-dimensional systems, so no transitions occur in the
high-noise limit. The details of this study will be published
elsewhere \cite{lop}.

We have shown that our driven sandpile model displays a variety
of steady states and many transitions between them. This
richness, coupled with its simplicity, makes our model a
promising one for the study of nonequilibrium phenomena in
driven systems. As noted above, it displays transitions similar
to those in other driven systems, e.g., our onset transitions
(Fig. 2) are like unpinning transitions in sliding CDW's \cite
{onu} or in pinned flux-lattice systems
\cite{flux}. It would be interesting to study whether this
similarity is merely superficial. There are some obvious ways in
which the CDW models are different from ours: (1) They exhibit
pinning because of quenched randomness but have no external
noise; and (2) no current flows in their pinned states. The
importance of these differences needs to be elucidated. In this
general context it is interesting to study the zero-current
limit of our model. We find that it {\em does not} show
conventional SOC when the pile is allowed to relax, after each
input of sand, to a completely quiescent state in which no
further transfers are possible (\`{a} la Bak, Tang, and
Wiesenfeld \cite{btw}).  The precise forms of the distribution
of avalanche sizes, etc., will be reported elsewhere \cite{lop}.

\acknowledgments

We thank CSIR and BRNS (India) for support, and the SERC
(IISc. Bangalore) for computational facilities.

\begin{figure}
\caption{The nonequilibrium phase diagram of our model in the
$\alpha-j_{\rm in}\/$ plane. Full (dashed) lines indicate
continuous (first-order) transitions. At low $\alpha\/$ there is
an infinity of first-order  boundaries; only the first five
are shown. The corners in the phase boundaries and
multicritical points are schematic; our data are too noisy to
resolve them.}
\end{figure}

\begin{figure}
\caption{A plot (full line) of $\sigma_{\rm  av}$ versus
$j_{\rm in}$.  Dashed lines are the asymptotes for large
and small $j_{\rm in}\/$. Arrows mark successive onsets (see
text). The inset shows a finite-size scaling plot for
$\delta\sigma N^{\beta/\nu}\/$ versus $NJ^{\nu}\/$ at
$\alpha=0.1\/$ and $j_{\rm in} \protect\gtrsim 0.5\/$ for $N=64(+),
128(\diamond), 256(\ast)\/$ and $512(\bigtriangleup)\/$ with
$\beta=0.5\/$ and $\nu=4\/$.}
\end{figure}

\begin{figure}
\caption{Equal-time height correlation length $\xi_{hh}$ and output
current autocorrelation time $\tau_{jj}$ versus $j_{\rm in}$ for
$N=32 (\circ), 64 (+), 128 (\diamond), 256 (\ast)\, \mbox{and}\,
512 (\bigtriangleup)\/$.  For $\xi_{hh}$ we do not plot points
in regions where $\xi_{hh} > N\/$.}
\end{figure}

\begin{figure}
\caption{A plot of $j_i\/$ versus $\sigma_i\/$ (Eq. (2)) for
$\alpha=0.1$ and $0.125\/$ and $\sigma_c = 10\/$.}
\end{figure}


\begin{references}
\bibitem{foot2} The time required to reach the steady state
increases as $N^2\/$; also it is high for $\alpha \simeq
0\/$ and $\alpha \simeq 0.5\/$ and seems to increase monotonically
with decreasing $j_{\rm in}\/$.
\end{references}
\end{document}